\documentstyle[12pt]{article}
\begin{document}

\begin{flushright}
HU-TFT-93-39
\end{flushright}

\large
\centerline{\bf Morphological Changes of Small Viscous}
\centerline{\bf Droplets under Spreading}
\vskip10mm

\normalsize

\centerline{ O. Ven\"al\"ainen$^1$,  T. Ala - Nissila$^{1,2}$,
 and  K. Kaski$^{1,2}$ }

\vskip6mm
\centerline{ $^1$Department of Electrical Engineering}
\centerline{ Tampere University of Technology}
\centerline{ P.O. Box 692}
\centerline{ FIN - 33101 Tampere}
\centerline{ Finland}

\vskip6mm
\centerline{ $^2$Research Institute for Theoretical Physics}
\centerline{ University of Helsinki}
\centerline{ P.O. Box 9 (Siltavuorenpenger 20 C)}
\centerline{ FIN - 00014 University of Helsinki}
\centerline{ Finland}
\vskip6mm

{\it Abstract.--}
Dynamics of spreading of viscous non - volatile fluid droplets
on surfaces is modelled using a solid - on - solid model,
which is studied with Monte Carlo simulations.
Tendency for dynamical layering and surface attraction are
in part embedded into the effective dynamics of the
model. This allows a description of the spreading process
with a single parameter, which strongly influences the morphology of
the droplets. The results qualitatively reproduce many experimentally
observed density profiles for polymeric fluids, including
rounded droplet shapes, and dynamical layering.


\pagebreak
\baselineskip 24pt
\parindent=0mm

{\it Introduction.--}
Experimental measurements on spreading profiles of tiny
droplets on surfaces \cite{HesN,HesP1,HesP2,Alb,Val,Fra,Leg,deG}
have recently been facilitated on molecular scales by highly developed
optical techniques. When layer
thicknesses become of the order of {\AA}ngstr\"oms, complicated droplet
morphologies emerge \cite{HesN,HesP1,HesP2,Alb,Val}
which cannot be accounted for by standard hydrodynamic theories
\cite{Leg,deG,Tan,Joa}. For example,
thickness profiles of tetrakis (2 - ethylhexoxy) - silane and
polydimethylsiloxane (PDMS) droplets on a silicon wafer
exhibit strikingly different shapes under spreading \cite{HesN}.
Tetrakis forms clearly observable dynamic
layering (Fig. 1(a)), while the spreading of PDMS proceeds
by a fast evolving precursor layer of one molecular thickness.
Comparisons between flatter PDMS and squalane droplets
reveal a ``Mexican hat'' shape for PDMS, i.e. a flat rounded shape
with a small central cap, while squalane develops a more uniform profile
approaching a Gaussian shape at late times \cite{HesP1,HesP2}.
Further experiments reveal that PDMS develops dynamical layering, too,
on a high - energy surface \cite{HesP2}. On the other hand, another recent
measurement of PDMS on silver produces a spherical cap (Fig. 1(b)),
with Gaussian tails at very low densities \cite{Alb}. More complicated
shapes are also possible \cite{Val,Fra}.

\vskip3mm
On the theoretical side, attempts have been made to formulate
analytic theories which could explain the experiments. In particular,
dynamical layering can be obtained from the layered flow model of
de Gennes and Cazabat \cite{Fra,deG2}, and from
the horizontal solid - on - solid (SOS) model \cite{Abr1,Abr2,Hei}.
In the latter case, the layer thickness is determined by the
range and the strength of the substrate potential,
which can be related to the Hamaker constant
of the substrate. A more microscopic view is
obtained from molecular dynamics simulations \cite{Nie1,Nie2,Yan}.
For small Lennard - Jones (LJ) particles, dynamical layering
tends to occur \cite{Nie1,Nie2}.
On the other hand in mixtures of ``solvent''
(LJ) particles and {\it flexible chains},
the chain structure of the molecules often leads to {\it intermixing}
of molecular layers and more
rounded droplet shapes, except on very strongly attractive substrates
where dynamical layering is characteristic \cite{Nie2}. These results
suggest a microscopic explanation for the role of the surface
potential and the molecular structure of the fluid on the morphology
of spreading microdroplets.

\vskip3mm
However, molecular dynamics is limited to relatively small systems,
and it would be of interest to further explore the role of the
surface and the effective fluid viscosity on droplet shapes on a more
coarse grained level \cite{deG2}.
In this letter we shall present
results of numerical simulations of an SOS model,
which has been modified to more realistically describe the
spreading of viscous fluids \cite{deG,deG2}.
We shall demonstrate that the model can reproduce both
stepped (layered) and more rounded droplet profiles just by
changing temperature, which in the model controls the
interaction strengths of the particles and the surface, and
mixing of layers.

\vskip6mm
{\it The model.--}
The model is given by the following discrete hamiltonian:

$$H (\lbrace h \rbrace)={{\tau }\over 2} \sum_{i,j}\min ( h_i,h_j )+
 \sum_j V(h_j) h_j
  ,\eqno (1)$$

where $\tau < 0$ is a nearest neighbour
interaction parameter in the horizontal direction, and
$h_j$ refers to the height of a surface column at $j$.
The heights measure the average concentration
and obey the SOS restriction \cite{Lea}. $V(h_j)$ is an attractive
substrate potential field at $j$. We shall always work
under complete wetting conditions in the canonical ensemble,
where the final equilibrium state of the system is
a non - volatile surface gas \cite{Leg}.

\vskip3mm
An SOS model with a short ranged surface attraction and
nearest neighbour diffusion dynamics was used in the spreading dynamics
study of Heini\"o {\it et al.} \cite{Hei} who obtained droplet shapes
with a single precursor layer, and a central cap.
They did not carry out a systematic study, however.
To remedy some of the shortcomings of the SOS model, a {\it horizontal}
SOS model was conceived to describe dynamical layering \cite{Abr1}.
In this work, we take a different approach. Physics describing the
spreading of viscous fluids and the effect of the surface potential
are in part embedded into the effective {\it dynamics} of the model.
This is done by using standard nearest neighbour diffusion
jumps with Metropolis Monte Carlo dynamics, but with two additional
features. First, particles are neither allowed to climb over height barriers
nor move down. In other words, a jump from a site $j$
to a neighbouring site $j^\prime$ is allowed only if
$h_j - h_{j^\prime} = 1$.
Second, when $h_j - h_{j^\prime} \ge  2$
there is an additional permissible jump
for the particle sitting at site $j$ at the height $h_{j^\prime}+1$
to move on top of the column at $j^{\prime}$.
This means that with probability
$p=\min \{ 1,e^{-\Delta H/kT} \}$
a transient non - SOS excitation (hole) can be created
into the $j$th column by a particle, which jumps on top of the neighbouring
column. The energy difference $\Delta H$ is calculated for this transient
state (ignoring the hole) \cite{weaseltalk}.
To obey the SOS restriction the hole left behind is immediately
filled by the column of particles above it, which are all lowered by a unit
step.

\vskip3mm
The motivation for these unusual dynamical rules follows from the
physics of viscous fluids on attractive surfaces
\cite{Leg,deG,deG2,Nie1,Nie2}. The first rule guarantees a well defined
diffusion length for the effective particles. It describes
horizontal flow of layers, and incorporates
a long ranged surface attraction, which here is independent of
the temperature. The second rule is similar to the
idea of an interlayer permeation flow of de Gennes and
Cazabat \cite{deG2}. Lowering the temperature reduces the
flow, and should promote separation of layers \cite{Nie2}.
By changing $\Delta H/kT$ we should then be able to qualitatively model
spreading under a variety of conditions.
We note that in the spirit of the model,
the effective particles should be regarded
as coarse grained layers of molecules in the horizontal direction.
Thus, all comparisons with experimental results are qualitative at best.

\vskip6mm
{\it Results.--}
First, we set
$\tau = V(h_j) = 0$ for all $h_j$, corresponding to infinite temperature.
Because of the effective dynamics, even this case is nontrivial.
Two sizes of three dimensional
droplets were studied, namely $10 \times 10 \times 11$
and $40 \times 40 \times 41$, where the last figure is the initial number
of vertical layers. At earlier stages of spreading
the droplets assume a rounded shape, with Gaussian tails.
They spread very fast in the
beginning due to less effective mutual blocking of the particles. At later
stages spreading becomes much slower, and
droplets assume a Gaussian shape, with
radius $r(t)\sim t^{0.5}$.
We note that in the case of real polymer chains,
late time spreading under submonolayer conditions may still show
deviations from Gaussian behavior due to interchain
blocking effects \cite{Alb,Hjelt}.

\vskip3mm
Next, we set $V(h_j = 0)=0$, and $V(h_j >0)
=\tau $. The initial sizes of the droplets were chosen to be
$10 \times 10 \times 11$ particles, with some test runs done
with larger systems as above. Since $V(h_j>0)$ is constant,
the initial height of 11 is sufficiently large.
First, we discuss ``high temperature'' simulations
with $\vert kT/\tau \vert =1$. The droplet starts out with a rounded
shape which becomes more Gaussian at later times.
In this case the dominant phenomenon in the morphology
is the formation of a small cap in the center of
the droplet above the first layer (the ``Mexican hat'' shape), see Fig. 2.
These profiles are qualitatively similar to those of
PDMS on silicon wafers \cite{HesN,HesP1}.
Lowering the  temperature to $\vert kT/\tau \vert =0.8$
shows a more clear formation of a shoulder
at $t=1800$ Monte Carlo steps per site (MCS/s) in Fig. 3.
At $t=2200$ MCS/s a small central cap can be seen.
At $t=2600$ MCS/s it has disappeared and the spreading has become very slow
with the droplet assuming a more rounded shape with Gaussian
tails at the edges. Diffusion at the centre of the droplet
is effectively blocked \cite{Alb}.

\vskip3mm
Next, we have lowered the temperature to $\vert kT/\tau \vert =0.6$. This
case produces impressive effects and clarifies the evolution of a drop
with an effectively higher viscosity, and small interlayer flow at the edges.
{\it Dynamical layering} is evident in Fig. 4. Up to 5 - 6 layers can
be easily seen with bigger droplets.
The edge of the core region and the surface gas is sharp.
The final profile at $t=5800$ MCS/s indicates a pancake like
shape with a remnant of the small central cap.
These results bear a qualitative resemblance to experiments of
tetrakis on silicon, see Fig. 1(a).

\vskip3mm
We have also studied the effect of setting
$\vert kT/\tau \vert =\vert kT/V(h_j) \vert =1$ for all $h_j$.
This corresponds to a stronger attraction on the surface level.
The results indicate increased tendency for dynamical layering
between the first and second layers.
Thus, increased surface attraction tends to enhance
layering which is in accordance with the conclusions of Ref. \cite{Nie2}.

\vskip3mm
An important question concerns the effect of
a height dependent attraction $V(h_j) = A/h_j^3$ \cite{deG2,Nie1,Nie2}.
For our model with modified dynamics,
using such a potential enhances layering somewhat, and
larger droplets produce ``stepped pyramid'' morphologies
as seen e.g. in Ref. \cite{Fra}. Details of these results will be
published elsewhere.

\vskip3mm
Finally, we have also done additional runs using Eq. (1) with
unrestricted diffusion dynamics
in analogy to Heini\"o {\it et al.} \cite{Hei}.
In this case with a constant attraction $V(h_j)=\tau$
no evidence of dynamical layering was found. With a height dependent
attraction, indications of layering can be detected
only if very large values of $A$ are used.
This demonstrates how our model with modified dynamics better captures
the essential features of spreading of viscous droplets.

\vskip3mm
{\it Summary and Discussion.--}
To summarize, we have developed a simple coarse grained SOS model for the
dynamics of droplet spreading on surfaces. In the model, the layering tendency
of the fluid molecules and the surface interaction
have been in part embedded into the effective dynamics, which allows
a single parameter description of viscous fluid spreading
under a variety of conditions. At ``higher'' temperatures, which here
correspond to weaker interactions
and stronger interlayer permeation flow at the droplet edges, the droplets
initially assume rounded shapes. At later stages of spreading
this shape develops towards a Gaussian, as interparticle blocking effects
become less important. Also, a small central cap appears.
When temperature is lowered, the rounded shapes become more persistent,
and indications of layer separation in the vicinity of the surface strenghten.
At even lower temperatures where
interactions and effective viscosity increase, and
permeation flow weakens, clear dynamical multilayering
takes place. Inclusion of a height dependent surface attraction
enhances the range of layering.
These results are in accordance with analytic models
\cite{deG2,Abr1}, and molecular dynamics simulations of
microdroplets \cite{Nie1,Nie2}.

\vskip3mm
Finally, we would like to comment the relation of our results to the
experimental data. Due to the effective coarse grained nature
of the model, detailed comparisons are not possible. However,
experiments indicate enhanced layering on high energy surfaces \cite{HesP1},
and for more viscous droplets \cite{HesP1,HesP2}, in analogy with what
has been presented here. More systematic experimental studies
would be most desireable.

\vskip1cm

Acknowledgements: We wish to thank R. Dickman, S. Herminghaus,
and J. A. Nieminen and
for useful discussions,
and the Academy of Finland for financial support. We also wish to thank
P. Leiderer for sending his data on the density profiles of PDMS spreading
on silver.

\pagebreak

\centerline{\bf Figure captions}
\vskip1cm
\baselineskip 16pt

Fig. 1.
(a) Experimental thickness profiles of PDMS spreading on a high
energy silicon surface \cite{HesP2}. Dynamical layering occurs at late
times.
(b) Thickness profiles of PDMS on a silver surface after 50, 80, 150, and
360 minutes.
\cite{Alb}. These profiles can be fitted by
a spherical cap, with Gaussian convolution at the edges.

\vskip 1cm

Fig. 2.
Time development of a spreading drop profile with
$\vert kT/\tau \vert=1$. In Figs. 2, 3, and 4, $V(h_j=0)=0$ and
$V(h_j>0)=\tau$. At late times, a
tiny central cap is formed. The curves correspond to 400, 1400, and 1600 MCS/s
in descending order. Results in this and the following figures are
averages over 500 runs, and units are arbitrary.
\vskip 1cm

Fig. 3.
Time development of a spreading drop with
$\vert kT/\tau \vert=0.8$.
A shoulder and a central cap are clearly visible, with a rounded
cap shape and a Gaussian foot at later times.
Curves correspond to 2000, 2200, and 2600 MCS/s.
\vskip 1cm

Fig. 4.
Time development of a spreading droplet with
$\vert kT/\tau \vert=0.6$.
Dynamical layering is evident at later times (cf. Fig. 1(a)).
Curves correspond to 1800, 2800, and 3800 MCS/s.
Spreading dynamics has also slowed down.

\pagebreak


\begin{thebibliography}{99}
\bibitem{weaseltalk} Calculating the final energy for the relaxed
                     SOS state makes essentially no difference in our
                     results.
\end{thebibliography}
\end{document}